\documentclass[acmsmall,screen,nonacm]{acmart}
\AtBeginDocument{}
\setcopyright{none}
\renewcommand\footnotetextcopyrightpermission[1]{}
\acmYear{2026}
\copyrightyear{2026}
\acmDOI{}
\acmISBN{}
\usepackage{amsmath}
\usepackage{booktabs}
\usepackage{tabularx}
\usepackage{subcaption}
\usepackage{listings}
\usepackage{algorithm}
\usepackage{algpseudocode}

\algrenewcommand\algorithmicrequire{\textbf{Input:}}
\algrenewcommand\algorithmicensure{\textbf{Output:}}
\usepackage{tikz}
\usetikzlibrary{arrows.meta,positioning}
\newcommand{\irene}{\textsc{Irene}}

\newcommand{\ket}[1]{\lvert #1\rangle}
\newcommand{\bra}[1]{\langle #1\rvert}
\newcommand{\ind}[1]{[\,#1\,]}
\newcommand{\EQ}{\textsc{Eq}}
\newcommand{\NEQ}{\textsc{Neq}}
\newcommand{\UNK}{\textsc{Unknown}}
\DeclareMathOperator{\Tr}{Tr}

\newtheorem{definition}{Definition}

\newcommand{\NumPairs}{1,982}
\newcommand{\NumDecided}{1,584}
\newcommand{\NumApproxEq}{127}
\newcommand{\DecisionRate}{79.92\%}
\newcommand{\MeanTime}{3.93}
\newcommand{\TimeLimit}{600}
\newcommand{\MemoryLimit}{6}
\newcommand{\NumCompilerBugs}{15}
\newcommand{\NumReportedBugs}{17}

\newcommand{\ArtifactURL}{https://github.com/WindOctober/Irene}

\begin{document}
\title[Structure-Preserving Quantum Equivalence Checking]{Irene: Equivalence Checking of Hybrid Quantum Programs via Structure-Preserving Symbolic Reduction}
\author{Jingyu Ke}
\affiliation{\institution{Shanghai Jiao Tong University}
  \city{Shanghai}
  \country{China}}
\email{windocotber@sjtu.edu.cn}

\author{Jingyang Li}
\affiliation{\institution{Shanghai Jiao Tong University}
  \city{Shanghai}
  \country{China}}
\email{lijjjjj@sjtu.edu.cn}

\author{Guoqiang Li}
\affiliation{\institution{Shanghai Jiao Tong University}
  \city{Shanghai}
  \country{China}}
\email{li.g@sjtu.edu.cn}
\begin{abstract}
Equivalence checking is essential to validating compiler transformations
of hybrid quantum programs, which combine quantum operations,
measurements, and classical control. Measurement-dependent control limits
the applicability of unitary reasoning, while dependencies between
classical outcomes and quantum operations can cause substantial growth
in intermediate symbolic states.
We present \irene{}, an equivalence-checking framework for bounded hybrid
quantum programs based on \emph{structure-preserving symbolic reduction}.
The framework progressively simplifies both sides of an equivalence
obligation through three levels of reasoning. At the gate level, algebraic
identities simplify unitary regions of the programs. At the hybrid path-sum
(HPS) level, reduced symbolic execution states are represented as typed
graphs, whose isomorphism certifies equivalence. For obligations that remain
unresolved, density kernels characterize the programs' transformations of
input density operators into observable outputs, allowing comparison even
when internal measurement histories differ. Residual differences between
the kernel coefficient expressions are encoded as SMT queries.
Simplifications established at each level are preserved in subsequent
reasoning, with explicit expansion restricted to obligations not discharged
by preceding reductions. A common set of symbolic reductions supports both HPS
and density-kernel reasoning: it operates on Boolean and arithmetic
expressions in factored form, eliminating reducible dependencies before
expanding residual sums, thereby limiting intermediate growth.
We implement the approach and evaluate \irene{} against five equivalence
checkers on \NumPairs{} program pairs from seven benchmark suites.
\irene{} solves \NumDecided{} pairs (\DecisionRate{}), compared with
57.52\% for MQT QCEC, the baseline with the highest aggregate coverage.
Its mean end-to-end time is \MeanTime{} seconds per solved pair.
Applied as an equivalence-checking oracle, \irene{} additionally identifies
\NumCompilerBugs{} previously unknown bugs in quantum compilers,
including Qiskit, Cirq, and PennyLane.
\end{abstract}
\maketitle
\section{Introduction}
\label{sec:intro}
As quantum computing advances, hybrid quantum--classical execution has
emerged as an important computing paradigm~\cite{preskill2018nisq}.
Integrating classical computation can
relax coherence-time requirements, while qubit reuse after measurement and
reset can reduce physical-qubit requirements~\cite{peruzzo2014vqe,decross2023reuse}.
Applications of such \emph{hybrid quantum programs} include molecular
ground-state energy estimation~\cite{peruzzo2014vqe} and combinatorial
optimization, including MaxCut~\cite{decross2023reuse}.
Deploying these applications on quantum hardware requires programs to
satisfy device-specific instruction-set and connectivity constraints.
For example, on connectivity-constrained devices, native two-qubit gates
can act only on directly connected physical qubits~\cite{li2019sabre}.
Quantum compilers therefore decompose gates, assign program qubits to
physical qubits, and route their interactions to produce executable
programs, while optimizing gate overhead and qubit
utilization~\cite{li2019sabre,hua2023caqr}.

However, quantum-specific semantics make these transformations error-prone.
For example, Giallar reports a Qiskit optimization that incorrectly merged
single-qubit gates with classical or quantum control
conditions~\cite{tao2022giallar}.
A valid gate identity is therefore insufficient unless the transformation
also respects the conditions under which the gates execute.
Furthermore, equality of measurement-outcome
distributions in a fixed basis is insufficient to establish program
equivalence:
$\ket{\pm}=(\ket{0}\pm\ket{1})/\sqrt{2}$ have identical computational-basis measurement
distributions, yet a subsequent Hadamard gate distinguishes
them~\cite{nielsen2010quantum}.
Measurement and feedback also induce correlations between classical
outcomes and the remaining quantum state. We therefore consider
\emph{observational equivalence}: two programs must produce the same
classical-output probabilities and associated quantum states on the selected
outputs for every input, including inputs entangled with an external
reference. Their internal measurement traces need not coincide.

Existing quantum program verification methods face a central scalability
challenge: controlling the growth of intermediate representations.
Even when the final relation between programs is simple, establishing it
can require substantially larger intermediate
representations~\cite{burgholzer2021advanced,chareton2026hps}.
Symbolic path-sum methods avoid explicit state enumeration, but
substitution and reduction can expand Boolean and phase expressions into
substantially larger algebraic
forms~\cite{amy2019pathsums,ricciardi2025sqbricks,huang2026quprs}.
In hybrid quantum programs, the reuse of measurement outcomes in
subsequent conditions and quantum operations introduces further dependencies
among these expressions~\cite{chareton2026hps,ricciardi2025sqbricks}.
Premature expansion can obscure cancellations and independent
subcomputations, increasing the cost of otherwise simple equivalence proofs.
Effective reduction must expose these simplifications while preserving
factored structure and the dependencies necessary for sound reasoning.

An additional challenge is to exploit equivalences that span measurement
and classical control.
Many efficient unitary equivalence checks rely on reversibility, which
measurement and reset invalidate in
general~\cite{burgholzer2022nonunitary}.
Equivalent programs may differ in their internal measurement outcomes and
branching structure; consequently, their equivalence need not admit a
one-to-one correspondence between execution branches~\cite{ricciardi2025sqbricks}.
Simplification restricted to unitary blocks may therefore fail to establish
equivalences arising from the combined effect of measurements and
conditional operations. Reasoning across these operations, however,
requires accounting for both
measurement-dependent control and correlations with other
qubits~\cite{amy2025relational,chareton2026hps}.
Consequently, checking a localized transformation can involve dependencies
extending well beyond the modified statements, making its cost sensitive
to the complexity of the surrounding program.

\paragraph{Our approach.}
We present \irene{}, a symbolic equivalence checker for
bounded hybrid quantum programs.
Our approach organizes \emph{structure-preserving symbolic reduction}
into three stages: gate-level reasoning, hybrid path-sum (HPS) reasoning,
and density-kernel comparison. Each stage simplifies the comparison and
seeks a certificate before invoking the next level of reasoning.
This progression exploits gate identities first, then symbolic
structure, and finally the programs' observable behavior.

First, gate-level reasoning simplifies unitary sequences using algebraic
identities. For eligible unitary pairs, \irene{} composes one program with
the inverse of the other and checks whether the composition is
observationally equivalent to $\mathsf{skip}$. Comparisons settled at this
stage require no subsequent HPS matching or density-kernel construction.

Second, HPS reasoning symbolically executes unresolved comparisons using
the hybrid path-sum semantics~\cite{chareton2026hps}, which represents
quantum amplitudes together with classical memory and measurement histories.
\irene{} represents Boolean expressions as XOR--AND graphs (XAGs),
retaining nested XOR and AND operations rather than distributing products
into algebraic normal form (ANF). Reduction operates directly on these graphs
to expose cancellations, eliminate summation variables, and identify
separable path sums. The reduced representations are then compared for
a sufficient structural certificate.

Finally, density-kernel comparison handles obligations not settled by HPS
reasoning. The kernels describe how input density operators determine the
selected classical and quantum outputs, accounting for quantum coherence
and hidden histories without requiring
matching internal decompositions. Graph-based expression reduction and
path-sum planning continue at this level to simplify
the kernel difference. Only residual sums require expansion, after which
SMT queries discharge the remaining coefficient-equality obligations.
Thus, structure-preserving reduction supports both symbolic levels, limiting
intermediate growth throughout the staged procedure.

\paragraph{Evaluation.}
We compare \irene{} with five equivalence checkers on \NumPairs{} program
pairs across seven benchmark suites. \irene{} solves
\NumDecided{} pairs (\DecisionRate{}), compared with 1,140 (57.52\%) for
MQT QCEC, the strongest baseline in aggregate coverage. Its mean
end-to-end time over solved pairs is \MeanTime{} seconds.
In ablations, disabling gate-level reasoning, feedback
summaries, expression simplification, or path-sum planning reduces coverage
by 111 to 540 pairs. Gate-level reasoning yields the largest net coverage
benefit among these components. Storage estimates that account for sharing
indicate that XAGs require at most half the explicit ANF storage for 145 of
1,549 program pairs with exact counts for both representations.
We further use \irene{} as an equivalence-checking oracle for LLM-assisted
fuzzing of optimization passes in Qiskit, Cirq, tket, PyZX, and PennyLane.
The LLM generates test programs tailored to the input requirements of each
pass, and \irene{} checks equivalence between the original and transformed
programs. This evaluation identifies \NumCompilerBugs{} previously unknown bugs.

\paragraph{Contributions.}
This paper makes the following contributions:
\begin{enumerate}
\item We propose a structure-preserving symbolic equivalence-checking
framework for bounded hybrid quantum programs that progresses from
gate-level reasoning through HPS structural proofs to density-kernel
comparison. Reduction at both symbolic levels retains factored expressions
and the dependencies required for sound reasoning,
simplifying comparisons before subsequent reasoning and residual expansion.
\item We implement a prototype, \irene{}, and evaluate it on \NumPairs{}
benchmark pairs. \irene{} solves \NumDecided{}
(\DecisionRate{}) with a mean end-to-end time of \MeanTime{} seconds per
solved pair, demonstrating the effectiveness and efficiency of our method.
\item We combine \irene{} with LLM-assisted fuzzing to uncover \NumCompilerBugs{}
previously unknown bugs in compilation and optimization passes across
Qiskit, Cirq, tket, PyZX, and PennyLane.
Our evaluation additionally identifies one implementation bug in each of
SQbricks and HQbricks~\cite{chareton2026hps}; both have been reported to the
respective developers.
\end{enumerate}

\section{Preliminaries}
\label{sec:semantics}
This section specifies the program syntax, symbolic execution model, and
notion of observational equivalence used throughout the paper.
\subsection{Hybrid Quantum Programs}
\label{sec:gates}
A hybrid quantum program operates on quantum registers and classical bits;
measurement produces classical outcomes that may determine subsequent
operations. We consider bounded hybrid quantum programs with the following
syntax:
\begin{equation}
\begin{aligned}
 P ::= {}& \mathsf{skip}\mid U(\mathbf q)\mid
 \mathsf{reset}\ q\mid c:=\mathsf{measure}\ q\\
 &{}\mid c:=e\mid P;P\mid
 \mathsf{if}\ e\ \mathsf{then}\ P\ \mathsf{else}\ P .
\end{aligned}
\label{eq:syntax}
\end{equation}
Here $e$ is a Boolean expression over classical bits, and $U$ is a unitary
gate. The grammar defines a finite core language: supported source-level
loops with statically determined iteration ranges are expanded into finite
sequences before symbolic execution.

An $n$-qubit register has state space
$\mathcal H_n=(\mathbb C^2)^{\otimes n}$ with computational basis
$\{\ket{\mathbf x}:\mathbf x\in\{0,1\}^n\}$, where
$\ket{\mathbf x}=\ket{x_1}\otimes\cdots\otimes\ket{x_n}
=\ket{x_1\cdots x_n}$~\cite{selinger2004qpl}.
We write $\bra{\psi}=\ket{\psi}^{\dagger}$, where $\dagger$ denotes
conjugate transpose, and $\mathcal L(\mathcal H_n)$ for the linear operators
on $\mathcal H_n$. A register state is a density operator in
$\mathcal D(\mathcal H_n)
=\{A\in\mathcal L(\mathcal H_n)\mid A\succeq0,\ \Tr(A)=1\}$,
where $A\succeq0$ denotes positive semidefiniteness.
Density operators represent populations, coherences, and correlations,
including entanglement. A pure state
$\ket{\psi}=\sum_{\mathbf x}\alpha_{\mathbf x}\ket{\mathbf x}$,
with $\alpha_{\mathbf x}\in\mathbb C$ and
$\sum_{\mathbf x}|\alpha_{\mathbf x}|^2=1$, has density operator
$\rho=\ket{\psi}\bra{\psi}$~\cite{nielsen2010quantum}.

A unitary gate satisfies $U^\dagger U=UU^\dagger=I$, with $I$ the identity,
and acts as $\ket{\psi}\mapsto U\ket{\psi}$ or
$\rho\mapsto U\rho U^\dagger$~\cite{selinger2004qpl}.
All local operators below are implicitly extended by the identity on the
remaining qubits. By linearity, gates are determined by their basis-state
actions; for $x,z\in\{0,1\}$,
\begin{equation}
 X\ket{x}=\ket{x\oplus1},\quad
 Z\ket{x}=(-1)^x\ket{x},\quad
 T\ket{x}=e^{i\pi x/4}\ket{x},\quad
 H\ket{x}=\frac1{\sqrt2}\sum_z(-1)^{xz}\ket{z}.
 \label{eq:gates}
\end{equation}
Computational-basis measurement uses $M_b=\ket{b}\bra{b}$ for
$b\in\{0,1\}$. It records $b$ classically with probability
$\Tr(M_b\rho M_b^\dagger)$ and unnormalized branch state
$M_b\rho M_b^\dagger$~\cite{born1926collisions,selinger2004qpl}.

Reset uses $\ket{0}\bra{0}$ and $\ket{0}\bra{1}$ to discard the previous
qubit state and prepare $\ket{0}$~\cite{nielsen2010quantum}.

\subsection{Hybrid Path-Sum Execution}
\label{sec:symbolic}
Let $\mathbf{x}\in\{0,1\}^n$ denote the symbolic computational-basis labels
of the $n$ input qubits, and let $N$ denote the total number of input and
auxiliary qubits. A \emph{path-sum} representation~\cite{amy2019pathsums}
is a tuple $S=(w,\mathbf y,\phi,O_Q)$ with denotation
\begin{equation}
 \ket{\psi_S(\mathbf{x})}
 =\sum_{\mathbf y\in\{0,1\}^m}w(\mathbf{x},\mathbf y)
 e^{2\pi i\phi(\mathbf{x},\mathbf y)}
 \ket{O_Q(\mathbf{x},\mathbf y)}.
 \label{eq:path-sum-state}
\end{equation}
Here $\mathbf y=(y_1,\ldots,y_m)$ is a vector of Boolean path variables bound by
the sum, $w:\{0,1\}^{n+m}\to\mathbb R$ is a real-valued
coefficient function, $\phi:\{0,1\}^{n+m}\to\mathbb R$ specifies the
phase in turns, interpreted modulo $1$, and $O_Q:\{0,1\}^{n+m}\to\{0,1\}^{N}$ is the \emph{quantum
output signature} of the register, with one coordinate per qubit: $O_Q(q)$
denotes the coordinate of qubit $q$, and $O_Q|_{W}$ the coordinates on a list
$W$ of qubits, in the order of $W$. The coefficient $w$ may be negative.
For fixed $\mathbf x$, each valuation of $\mathbf y$ identifies a
\emph{path} with amplitude $w(\mathbf x,\mathbf y)e^{2\pi i\phi(\mathbf x,\mathbf y)}$.
Amplitudes with the same output basis state add coherently.
The sum is represented symbolically without enumerating its $2^m$ paths.
For example, the Hadamard gate introduces a path variable $y$ with
$w(x,y)=1/\sqrt2$,
$\phi(x,y)=xy/2\pmod 1$, and $O_Q(x,y)=y$.

Following the hybrid path-sum (HPS) execution model~\cite{chareton2026hps},
\irene{} represents a complete symbolic execution state as a collection
$\Sigma$ of hybrid path-sum components. Each component
$D=(G,w,\mathbf y,\phi,(O_C,O_Q),\mathsf{H})$ extends the path-sum with
a Boolean guard $G$, symbolic classical memory $O_C$, and a history
$\mathsf{H}$ of measurement and discarded-value events.
A component denotes a guarded sum over paths, not an individual path.
Classical bits have fixed initial values; all symbolic expressions depend
on $(\mathbf x,\mathbf y)$.

For each classical-memory valuation $\mathbf c$ and history $\mathbf h$,
the denotation of $\Sigma$ is the unnormalized quantum state vector
\begin{equation}
 \ket{\psi_{\Sigma,\mathbf c,\mathbf h}(\mathbf x)}
 =\sum_{D\in\Sigma}\sum_{\mathbf y\in\{0,1\}^{m_D}}
 \ind{G_D}\ind{O_{C,D}=\mathbf c}\ind{\mathsf{H}_D=\mathbf h}
 w_D e^{2\pi i\phi_D}\ket{O_{Q,D}}.
 \label{eq:hps-state}
\end{equation}
Here $m_D$ counts the bound path variables in $D$, and $[G_D]$ equals
$1$ if $G_D$ holds and $0$ otherwise. All component expressions are evaluated at
$(\mathbf x,\mathbf y)$; these arguments are omitted for readability.
For each fixed pair $(\mathbf c,\mathbf h)$, paths from all contributing
components combine coherently. Contributions from distinct histories are
combined at the density-operator level. For a unitary program without
classical memory or history, this denotation reduces to
Eq.~\eqref{eq:path-sum-state}.

The initial collection $\Sigma$ contains one component: input qubits are
represented by $\mathbf x$, initialized auxiliary qubits by constants,
and the remaining fields satisfy $G=\top$, $w\equiv1$, and $\phi=0$,
with no bound paths or recorded histories.

Symbolic execution applies each statement componentwise to $\Sigma$,
collecting the results into a successor state; $\mathsf{skip}$ leaves
$\Sigma$ unchanged. A unitary gate updates $w$, $\phi$, and $O_Q$ by its
basis-state action, introducing fresh path variables when needed.
Measurement $c:=\mathsf{measure}\ q$ assigns $O_Q(q)$ to $O_C(c)$ and
records the same expression in $\mathsf{H}$. Distinct outcomes remain represented
symbolically rather than being enumerated. Reset records the discarded basis value in
$\mathsf{H}$ and sets $O_Q(q)=0$.
Classical assignment evaluates $e$ in $O_C$ and updates $O_C(c)$ without
changing $\mathsf{H}$. A conditional evaluates its condition in each
component's $O_C$ to obtain $b$, then executes the branches under
$G\land b$ and $G\land\neg b$. Sequential composition executes the next
statement on the resulting collection.

\begin{figure}[t]
 \centering
 \begin{subfigure}[t]{0.44\linewidth}
  \centering
  \includegraphics[width=\linewidth]{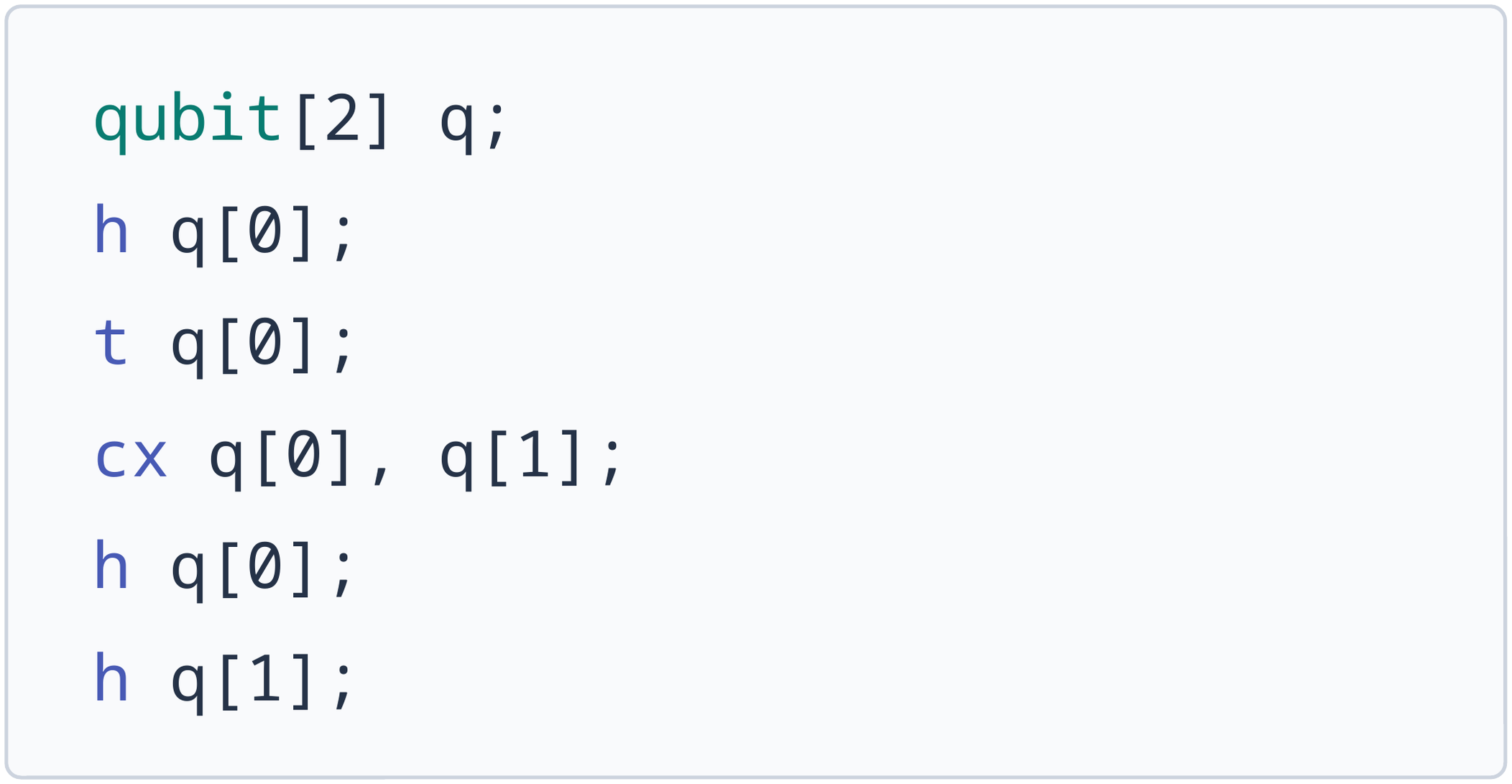}
  \caption{Unitary quantum program.}
  \label{fig:openqasm-unitary}
 \end{subfigure}\hspace{0.04\linewidth}
 \begin{subfigure}[t]{0.44\linewidth}
  \centering
  \includegraphics[width=\linewidth]{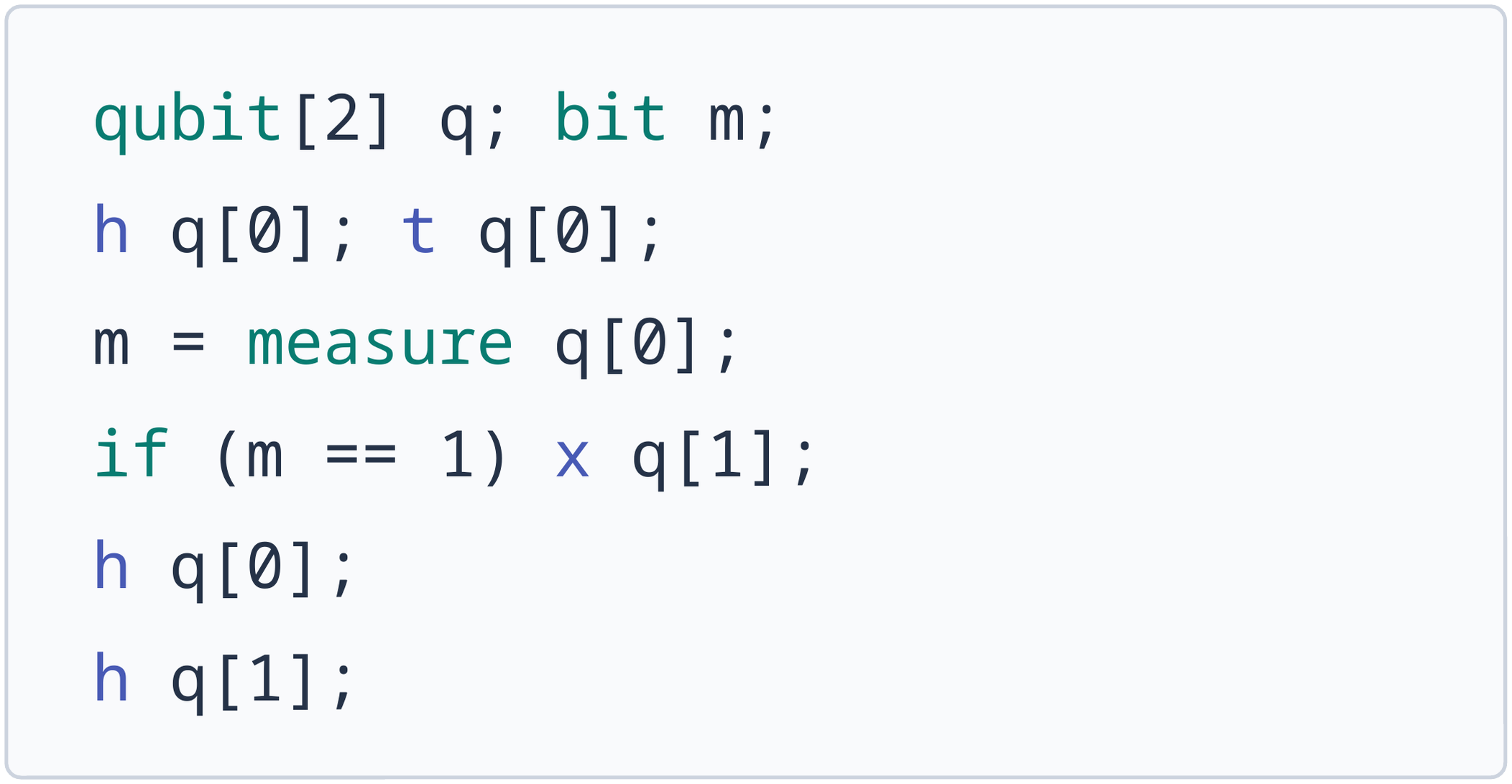}
  \caption{Hybrid quantum program.}
  \label{fig:openqasm-hybrid}
 \end{subfigure}
 \caption{OpenQASM examples for symbolic execution.}
 \label{fig:openqasm-examples}
 \Description{Two aligned six-line OpenQASM snippets on two qubits. Both
 apply H and T to the first qubit and end with H on each qubit. The left
 uses a CNOT; the right measures its control into m and conditionally
 applies X to the second qubit.}
\end{figure}

Figure~\ref{fig:openqasm-examples} compares unitary and hybrid programs with
the same Hadamard and $T$ gates. Figure~\ref{fig:openqasm-hybrid} replaces
the CNOT of Figure~\ref{fig:openqasm-unitary} with a measurement followed
by a classically controlled $X$.

\begin{example}[Unitary and hybrid symbolic execution]
\label{ex:fig1-execution}
Consider the programs in Figure~\ref{fig:openqasm-examples}, each initialized
to $\ket{00}$, with qubit order \texttt{q[0]}, \texttt{q[1]}.
For the unitary program, let $\mathbf y=(r,s,t)$ be the path variables
introduced by the three Hadamard gates, in execution order. Its final
symbolic state is $\Sigma_U=\{D_U\}$, whose sole component has
$G=\top$, $O_C=\varnothing$, $\mathsf{H}=\epsilon$ (the empty history),
$w=1/(2\sqrt2)$, $\phi=r/8+rs/2+rt/2$, and $O_Q=(s,t)$.
Each Hadamard contributes a scalar factor $1/\sqrt2$. The $T$ gate contributes $r/8$,
and the CNOT sets the second qubit's output coordinate to $r$. The final
Hadamards therefore contribute $rs/2$ and $rt/2$, replacing the two
coordinates by $s$ and $t$.

In the hybrid program, measurement stores $r$ in \texttt{m} and records
$\mathsf{H}=(r)$ without splitting the component. The conditional produces a state
$\Sigma_{\mathsf{if}}=\{D_0,D_1\}$: for $b\in\{0,1\}$, $D_b$ has
guard $r=b$ and quantum output signature $O_Q=(r,b)$; both components retain
$w=1/\sqrt2$, $\phi=r/8$, $O_C(\texttt{m})=r$, and $\mathsf{H}=(r)$.
After the final Hadamards, the phase of component $D_b$ is
$r/8+rs/2+bt/2$, which coincides with the unitary phase whenever its guard
holds. The two components therefore admit a single-component representation
with the same $G$, $\mathbf y$, $w$, $\phi$, and $O_Q$ as $D_U$,
but with $O_C(\texttt{m})=r$ and $\mathsf{H}=(r)$.
For each output $(s,t)$, the unitary program combines the amplitudes for
both values of $r$ coherently. In the hybrid program, the corresponding
paths have distinct measurement histories and do not interfere, even when
\texttt{m} is not observed. The probability of $(s,t)$ is therefore the sum of the squared
magnitudes of the amplitudes associated with the two histories, rather
than the squared magnitude of their sum.
\end{example}

\subsection{Observable Outputs and Equivalence}\label{sec:observable}
Let $P,Q$ be programs equipped with an input/output correspondence $\iota$.
The correspondence consists of a bijection between their $n$ input qubits
and ordered selections of $\ell$ output qubits and $k$ output classical
bits in each program, paired position-wise.
As in Section~\ref{sec:symbolic}, symbolic execution begins with input
labels $\mathbf x$ and fixed initial values for auxiliary qubits and
classical bits; external classical inputs are excluded. The internal
registers of $P$ and $Q$ need not coincide.

Observation retains only the selected classical bits and traces out
unselected qubits~\cite{ricciardi2025sqbricks}. Let $\mathbf v$ and
$\mathbf s$ be the selected output qubits and output classical bits of a
program, ordered as in $\iota$, and let $\overline{\mathbf v}$ be its remaining
qubits in a fixed order. The projection of a component
$D=(G,w,\mathbf y,\phi,(O_C,O_Q),\mathsf{H})$ is defined by
\[
 D|_{\iota}
 =\bigl(G,\;w,\;\mathbf y,\;\phi,\;
   (O_C|_{\mathbf s},\,O_Q|_{\mathbf v}),\;
   \mathsf{H}\cdot O_Q|_{\overline{\mathbf v}}\bigr),
\]
where $O_Q|_{\mathbf v}$ and $O_C|_{\mathbf s}$ denote the ordered
restrictions to the selected quantum and classical outputs, respectively.
Concatenation, denoted by $\cdot$, appends the discarded coordinates
$O_Q|_{\overline{\mathbf v}}$ to the recorded history.
Two paths have the same extended history if and only if they agree on
both the recorded events and the discarded output coordinates. For each
fixed recorded history, summing outer products over discarded labels
implements the partial trace: for retained labels $a,a'$ and
discarded labels $z,z'$,
$\Tr_{\overline{\mathbf v}}(\ket{a,z}\bra{a',z'})=\ind{z=z'}\ket{a}\bra{a'}$.
Hence, coherence is retained only between paths with the same recorded
history and discarded coordinates; distinct histories contribute additively
at the density-operator level. The classical memory is determined by its
initialization and the recorded history. Omitting an unselected classical
bit therefore does not identify distinct histories.

Let $\Sigma_P$ denote the final symbolic state of $P$ after output projection.
Equation~\eqref{eq:hps-state} assigns a vector
$\ket{\psi_{\Sigma_P,\mathbf c,\mathbf h}(\mathbf x)}$ to each selected
classical output $\mathbf c\in\{0,1\}^k$ and hidden history $\mathbf h$.
For an input density operator $\rho\in\mathcal D(\mathcal H_n)$, linear
extension to the outer-product basis yields the unnormalized output
density operator
\begin{equation}
 \rho_{P,\mathbf c}
 =\sum_{\mathbf x,\mathbf x'}
 \bra{\mathbf x}\rho\ket{\mathbf x'}
 \sum_{\mathbf h}
 \ket{\psi_{\Sigma_P,\mathbf c,\mathbf h}(\mathbf x)}
 \bra{\psi_{\Sigma_P,\mathbf c,\mathbf h}(\mathbf x')}.
 \label{eq:observable}
\end{equation}
Here $\mathbf x,\mathbf x'\in\{0,1\}^n$ independently index the input matrix
entries. With $K_{\mathbf c,\mathbf h}\ket{\mathbf x}=
\ket{\psi_{\Sigma_P,\mathbf c,\mathbf h}(\mathbf x)}$, the same operator is
$\sum_{\mathbf h}K_{\mathbf c,\mathbf h}\rho K_{\mathbf c,\mathbf h}^{\dagger}$,
which is positive semidefinite for every density operator $\rho$; only its
individual cross-input terms $\mathbf x\neq\mathbf x'$ need not be.
The trace $\Tr(\rho_{P,\mathbf c})$ is the probability of classical
output $\mathbf c$; normalizing $\rho_{P,\mathbf c}$ yields the corresponding
conditional quantum state whenever this probability is nonzero.
For terminating programs without
postselection, $\sum_{\mathbf c}\Tr(\rho_{P,\mathbf c})=1$.

\begin{definition}[Observational equivalence]
\label{def:obs-eq}
Fix an input/output correspondence $\iota$ and initial values for the
auxiliary qubits and classical bits of $P$ and $Q$.
The programs $P$ and $Q$ are \emph{observationally equivalent}, written
$P\equiv Q$, if and only if $\rho_{P,\mathbf c}=\rho_{Q,\mathbf c}$
for every input density operator $\rho\in\mathcal D(\mathcal H_n)$
and every classical output valuation $\mathbf c\in\{0,1\}^k$.
\end{definition}
This definition requires equality of both classical-output probabilities
and the associated quantum states, including their coherences. It also
preserves correlations with an untouched entangled reference, without
requiring identical internal histories or component decompositions.
Observational equivalence is the semantic target of \irene{}.
Section~\ref{sec:method} derives an equivalent, basis-indexed kernel
characterization from Eq.~\eqref{eq:observable} by linearity and uses it
to discharge residual equivalence obligations. Comparisons for which no
conclusive result is established are reported as \UNK{}.

\section{Structure-Preserving Equivalence Checking}\label{sec:method}
We now present \irene{}'s procedure for checking the observational
equivalence defined in Section~\ref{sec:semantics}.
The procedure progressively simplifies the two programs at the gate, HPS,
and density-kernel levels, preserving the structure needed for subsequent
reasoning and restricting expansion to unresolved obligations.

\subsection{Overview}\label{sec:overview}
Given programs $P,Q$, an input/output correspondence $\iota$, and a specified
initialization, \irene{} checks their observational equivalence.
Algorithm~\ref{alg:checker} organizes this reasoning
into three stages.

First, gate-level reasoning simplifies unitary regions using algebraic
identities. For unitary pairs satisfying the input--output correspondence
conditions in Section~\ref{sec:gate}, it composes one program with the inverse
of the other and checks whether the composition is observationally
equivalent to $\mathsf{skip}$.
Reducing the composition to the identity proves equivalence; a trace-based
criterion can establish either equivalence or inequivalence.

Second, HPS reasoning symbolically executes the program pair obtained after
unitary simplification and projects the outputs according to $\iota$,
producing HPS collections as described in
Section~\ref{sec:semantics}. Symbolic reduction simplifies the resulting
components while preserving factored expressions and the dependencies
required for sound reasoning.
A structural match between the resulting collections proves equivalence:
their typed HPS graphs must be isomorphic, allowing component reordering
and consistent renaming of bound path variables (Section~\ref{sec:pathwise}).

Finally, \irene{} compares the density kernels derived from
Eq.~\eqref{eq:observable}, which describe each program's action on input
density operators. This comparison accounts for quantum coherence and
hidden histories without requiring a structural match between programs.
Symbolic reduction simplifies the kernel difference before residual sums
are evaluated and their coefficient expressions compared through SMT
queries (Section~\ref{sec:decision}). Thus, the comparison moves from
gate identities to symbolic structure and finally to observable behavior.
Graph-based expression reduction and path-sum planning support
both symbolic levels (Section~\ref{sec:cooperative}), retaining factored
expressions and sums throughout this progression.

\begin{algorithm}[t]
\caption{\textsc{Check}: three-stage observational equivalence checking}
\label{alg:checker}
\begin{algorithmic}[1]
\Require Programs $P,Q$, validated correspondence $\iota$ and initialization
\Ensure $\EQ{}$, $\NEQ{}$, or $\UNK{}$
\State $P_0\gets\Call{GateReduce}{P}$\Comment{\textbf{Stage 1: gate-level reasoning}}
\State $Q_0\gets\Call{GateReduce}{Q}$
\If{$\Call{IsUnitary}{P_0,Q_0,\iota}$}
  \State $M\gets\Call{GateReduce}{P_0;Q_0^{-1}}$
  \State $r\gets\Call{TraceCheck}{M}$
  \State \textbf{if} $r\ne\UNK{}$ \textbf{then} \Return $r$
\EndIf
\State $\Sigma_P\gets\Call{Execute}{P_0,\iota}$\Comment{\textbf{Stage 2: HPS structural proof}}
\State $\Sigma_Q\gets\Call{Execute}{Q_0,\iota}$
\State $(\Sigma_P,\Sigma_Q)\gets\Call{ReduceHPS}{\Sigma_P,\Sigma_Q}$\label{alg:reduce-hps}
\State \textbf{if} $\Call{IsIsomorphic}{\Sigma_P,\Sigma_Q}$ \textbf{then} \Return $\EQ{}$
\State $\Delta\gets\Call{KernelDifference}{\Sigma_P,\Sigma_Q}$\Comment{\textbf{Stage 3: density-kernel checking}}
\State $\Delta\gets\Call{ReduceKernel}{\Delta}$\label{alg:reduce-kernel}
\State \textbf{if} $\Delta=0$ \textbf{then} \Return $\EQ{}$
\State $\Phi\gets\Call{EncodeFormula}{\Delta}$
\State \Return $\Call{Solve}{\Phi}$\Comment{SMT solving}
\end{algorithmic}
\end{algorithm}

The algorithm returns \EQ{} or \NEQ{} only when the corresponding relation
is established; otherwise it returns \UNK{}. Failure of a sufficient
certificate does not establish inequivalence. In particular, an
inconclusive trace check or structural comparison leaves the obligation
for subsequent reasoning. \textsc{ReduceHPS} (line~\ref{alg:reduce-hps})
and \textsc{ReduceKernel} (line~\ref{alg:reduce-kernel}) apply the common
reductions described in Section~\ref{sec:cooperative}.

\subsection{Gate-Level Reasoning}\label{sec:gate}
In the syntax of Eq.~\eqref{eq:syntax}, $U(\mathbf q)$ applies a unitary
gate $U$ to an ordered tuple of qubits $\mathbf q$. Gate-level reduction
acts on sequences of such applications within a single control-flow branch,
without crossing measurements or resets. Table~\ref{tab:gate-reduce}
presents representative identities for \textsc{GateReduce} in
Algorithm~\ref{alg:checker}. Quantum operands are omitted: each identity is
instantiated on the same ordered operands, preserving control and target
positions. For example, $HXH\to Z$ denotes the program rewrite
$H(q);X(q);H(q)\to Z(q)$, with all gates acting on the same qubit $q$.

Each rewrite requires matching execution conditions. A nonadjacent gate
may be brought into the sequence only if it commutes with every intervening
operation; rewrites do not cross control-flow boundaries. These identities
preserve the operator, including its phase.

\begin{table}[t]
\caption{Representative identities for \textsc{GateReduce}, where
$a\in\{x,y,z\}$ and $k\in\mathbb Z$. Rules require matching operands
and execution conditions; reordering additionally requires commutation.}
\label{tab:gate-reduce}
\centering
\small
\renewcommand{\arraystretch}{1.2}
\begin{tabularx}{\linewidth}{@{}c@{\enspace}>{\centering\arraybackslash}X@{\qquad}c@{\enspace}>{\centering\arraybackslash}X@{}}
\toprule
\textbf{Rule} & \textbf{Name} & \textbf{Rule} & \textbf{Name}\\
\midrule
$U^\dagger U\to I$ & Inverse cancellation &
$HH\to I$ & Hadamard cancellation\\
\addlinespace[2pt]
$\mathrm P(2k\pi)\to I$ & Phase periodicity &
$R_a(4k\pi)\to I$ & Rotation periodicity\\
\addlinespace[2pt]
$\mathrm P(\alpha)\mathrm P(\beta)\to\mathrm P(\alpha+\beta)$ & Phase fusion &
$R_a(\alpha)R_a(\beta)\to R_a(\alpha+\beta)$ & Rotation fusion\\
\addlinespace[2pt]
$HXH\to Z$ & $X$-to-$Z$ conjugation &
$HZH\to X$ & $Z$-to-$X$ conjugation\\
\bottomrule
\end{tabularx}
\end{table}

\begin{definition}[Unitary program and inverse]\label{def:unitary-program}
A \emph{unitary program} is a finite sequence of gate applications
$P=U_1(\mathbf q_1);\cdots;U_k(\mathbf q_k)$, with $\mathsf{skip}$
denoting the empty sequence. Its \emph{inverse} reverses the sequence and
replaces each gate by its adjoint:
$P^{-1}=U_k^\dagger(\mathbf q_k);\cdots;U_1^\dagger(\mathbf q_1)$,
with $\mathsf{skip}^{-1}=\mathsf{skip}$.
Writing $U_P$ for the induced operator on the full register, we have
$U_{P^{-1}}=U_P^\dagger$.
\end{definition}

The predicate \textsc{IsUnitary} in Algorithm~\ref{alg:checker} holds for
$(P,Q,\iota)$ when both programs are unitary on $n$ qubits and $\iota$
pairs all quantum inputs and outputs positionwise. Every qubit is an
arbitrary input and an observed output; there are no classical outputs.

\paragraph{Trace-based comparison.}
For such a pair, let $M=P;Q^{-1}$. Gate reduction on $M$ exposes
cancellations between the programs; reducing it to $\mathsf{skip}$ proves
equivalence. More generally, the standard trace
criterion~\cite{sander2025mpo} gives
\begin{equation}
 P\equiv Q
 \quad\Longleftrightarrow\quad
 M\equiv\mathsf{skip}
 \quad\Longleftrightarrow\quad
 |\Tr(U_M)|=2^n.
 \label{eq:unitary-miter}
\end{equation}
Accordingly, \textsc{TraceCheck} returns \EQ{} when the trace modulus is
certified to equal $2^n$, \NEQ{} when it is certified to be strictly smaller,
and \UNK{} when neither condition can be established.

\subsection{HPS Structural Certificates}\label{sec:representation}
\label{sec:pathwise}
HPS reasoning compares symbolic states with respect to the outputs selected
by $\iota$. Following Section~\ref{sec:observable}, each component is
projected as $D|_\iota$: selected outputs are retained, and discarded
quantum outputs are appended to $\mathsf{H}$ to account for the partial trace.
Let $\Sigma_P,\Sigma_Q$ denote the resulting collections.
This stage reduces their symbolic structure and uses isomorphism
between the resulting typed HPS graphs as a sufficient
certificate of program equivalence, without enumerating paths.

\paragraph{HPS expression representation.}
For $D=(G,w,\mathbf y,\phi,(O_C,O_Q),\mathsf{H})$, we represent its
fields by typed expressions over input variables $x_i$ and bound path
variables $y_j$. Boolean terms $B$ and real-valued weight terms $W$ have
the syntax
\[
 \begin{aligned}
 B&::=0\mid1\mid x_i\mid y_j\mid B\oplus B\mid B\land B,\\
 W&::=a\mid f(W_1,\ldots,W_s)\mid\mathsf{ite}(B,W_1,W_0).
 \end{aligned}
\]
Here $a\in\mathbb R$, and $f$ is a real-valued arithmetic operator of
arity $s$, applied within its domain. The conditional
$\mathsf{ite}(B,W_1,W_0)$ evaluates to $W_1$ when $B=1$ and to $W_0$
otherwise. Terms denote functions of $(\mathbf x,\mathbf y)$ as in
Section~\ref{sec:symbolic}. The guard $G$ is a Boolean term and the weight
$w$ a weight term. The phase is a weighted sum of Boolean terms:
\begin{equation}
 \phi=\sum_j a_j B_j(\mathbf x,\mathbf y)\pmod 1,
 \qquad B_j:\{0,1\}^{n+m_D}\to\{0,1\}.
 \label{eq:selector-phase}
\end{equation}
Here $a_j\in\mathbb R$ is a phase coefficient. The remaining fields are
ordered tuples of Boolean terms or labelled history events:
\[
 \begin{aligned}
 O_C&=(B^C_1,\ldots,B^C_k),\qquad
 O_Q=(B^Q_1,\ldots,B^Q_{\ell}),\\
 \mathsf{H}&=(\mathsf{event}_{\lambda_1}(B^{\mathsf{H}}_1),\ldots,
               \mathsf{event}_{\lambda_t}(B^{\mathsf{H}}_t)).
 \end{aligned}
\]
An event label $\lambda_i$ records a measurement or discard;
its Boolean child gives the recorded value.

Each expression has a syntax tree: $\kappa(e_1,\ldots,e_s)$ has a root
labelled $\kappa$ with the trees of $e_1,\ldots,e_s$ as children;
constants and variables are leaves. Tuple constructors preserve entry order. A phase-sum
node has weighted terms as children; each term carries its coefficient as
a label and has one Boolean child.
Boolean terms $B$, including conditions in $W$ and selectors in $\phi$,
are represented as XOR--AND graphs (XAGs), whose internal nodes are
$\oplus$ and $\land$ and whose leaves are constants and variables.
This representation retains nested Boolean operations without expanding
products into ANF.

\begin{definition}[Typed HPS graph]\label{def:hps-graph}
For a collection $\Sigma$, the graph $\mathfrak G(\Sigma)$ is a rooted,
node- and edge-labelled directed multigraph with a collection root $r$
and one component node $d_D$ per occurrence of $D$ in $\Sigma$.
The remaining nodes come from the syntax trees of each component's fields,
with identical Boolean terms shared within that component and one node
for every declared path variable, including unused ones.
Write $v_{D,e}$ for the root representing an expression occurrence $e$;
identical Boolean terms therefore have the same root.
The structural edges are
\[
 \begin{array}{@{}>{\displaystyle}c@{\qquad}l@{}}
 r\xrightarrow{\mathsf{member}}d_D,\quad
 d_D\xrightarrow{\mathsf{bind}}v_{D,y_i}
 &(1\le i\le m_D),\\
 d_D\xrightarrow{F}v_{D,F_D}
 &\bigl(F\in\{G,w,\phi,O_C,O_Q,\mathsf{H}\}\bigr).
 \end{array}
\]
For $e=\kappa(e_1,\ldots,e_s)$, expression edges
$v_{D,e}\xrightarrow{j}v_{D,e_j}$ encode operand positions in their
labels. Thus, for $W=\mathsf{ite}(B,W_1,W_0)$, the edges from $v_{D,W}$
to $v_{D,B}$, $v_{D,W_1}$, and $v_{D,W_0}$ carry labels $1$, $2$, and
$3$, respectively. A phase-term node for $a_jB_j$ carries the coefficient
$a_j$ modulo $1$ and connects to $v_{D,B_j}$ by an edge labelled $1$.
Edges from $\oplus$, $\land$, and phase-sum nodes instead share the
label $\mathsf{arg}$, with parallel edges retaining multiplicity.
Node labels preserve expression types and constructor labels, including
constant values and event labels $\lambda_i$.
Input-variable labels are fixed by $\iota$; bound-path nodes omit variable
names and are scoped by $\mathsf{bind}$ edges.
\end{definition}
Intuitively, label-preserving isomorphism matches components up to permutation
and consistent bijective renaming of their bound paths. It fixes input
coordinates and ordered operands, including output positions and history
event order, while allowing commutative operands to be reordered.
Matched expressions thus have the same denotation; component permutation
and path renaming only reindex the sums in Eq.~\eqref{eq:hps-state}. Hence,
$\mathfrak G(\Sigma_P)\cong\mathfrak G(\Sigma_Q)$ implies $P\equiv Q$.

\subsection{Density-Kernel Comparison}\label{sec:decision}
For equivalence obligations not discharged by HPS structural certificates,
\irene{} compares
the linear output maps of Eq.~\eqref{eq:observable} through their matrix
coefficients in the input and output matrix-unit bases. We call these
coefficient functions \emph{density kernels}.

\paragraph{Observable kernel.}
Let $\Sigma_P$ be the projected HPS collection of $P$. For classical output
$\mathbf c\in\{0,1\}^k$, quantum output indices
$\mathbf o,\mathbf o'\in\{0,1\}^{\ell}$, and input indices
$\mathbf x,\mathbf x'\in\{0,1\}^n$, the kernel is the
$(\mathbf o,\mathbf o')$ matrix entry of the inner history sum in
Eq.~\eqref{eq:observable}:
\begin{equation}
 \mathcal K_P(\mathbf c,\mathbf o,\mathbf o';\mathbf x,\mathbf x')
 =\sum_{\mathbf h}
 \langle\mathbf o\mid\psi_{\Sigma_P,\mathbf c,\mathbf h}(\mathbf x)\rangle
 \langle\psi_{\Sigma_P,\mathbf c,\mathbf h}(\mathbf x')\mid\mathbf o'\rangle.
 \label{eq:kernel}
\end{equation}
Thus, $\mathcal K_P(\mathbf c,\mathbf o,\mathbf o';\mathbf x,\mathbf x')$
is the $(\mathbf o,\mathbf o')$ entry obtained by applying the linear
output map to $\ket{\mathbf x}\bra{\mathbf x'}$. It specifies how each
input matrix entry contributes to an output entry, including the
off-diagonal entries that represent coherence. The matrix units are an
operator basis, not necessarily physical input states.

The sum over $\mathbf h$ combines distinct hidden histories at the
operator level, as in Section~\ref{sec:observable}, rather than adding
their amplitudes coherently. Within each history, the amplitude and its
conjugate use separate bound path variables, retaining all coherent
cross-component terms. Histories are summed out separately for each
program, so they need not correspond between $P$ and $Q$.

For an arbitrary input $\rho$, each output entry is therefore a linear
combination of input entries weighted by the kernel:
\[
 \bra{\mathbf o}\rho_{P,\mathbf c}\ket{\mathbf o'}
 =\sum_{\mathbf x,\mathbf x'}\bra{\mathbf x}\rho\ket{\mathbf x'}
 \mathcal K_P(\mathbf c,\mathbf o,\mathbf o';\mathbf x,\mathbf x').
\]
If the kernels agree, this identity gives equal output matrix entries
for every input density operator, hence $P\equiv Q$.
Conversely, $P\equiv Q$ means that, for each $\mathbf c$, the two linear
output maps agree on all density operators. Since these span
$\mathcal L(\mathcal H_n)$ over $\mathbb C$, the maps also agree on every
matrix unit $\ket{\mathbf x}\bra{\mathbf x'}$, giving equal kernel
entries. Thus,
\begin{equation}
 P\equiv Q\quad\Longleftrightarrow\quad
 \forall\mathbf u.\ \mathcal K_P(\mathbf u)=\mathcal K_Q(\mathbf u),
 \qquad \mathbf u=(\mathbf c,\mathbf o,\mathbf o',\mathbf x,\mathbf x').
 \label{eq:kernel-equiv}
\end{equation}
The independent input indices retain coherence: identity and $Z$ agree
on computational-basis input density operators but differ on
$\ket{+}\bra{+}$.

Since $\mathbf u\in\{0,1\}^{k+2\ell+2n}$, Eq.~\eqref{eq:kernel-equiv}
replaces equality over all input density operators with pointwise equality
of two functions on a finite Boolean domain. These functions are compared
symbolically through reduction and SMT queries, without enumerating all
index tuples or introducing symbolic input density matrices.

\paragraph{Residual coefficient normalization.}
Substituting the HPS amplitudes of Eq.~\eqref{eq:hps-state} into
Eq.~\eqref{eq:kernel} makes the kernel's algebraic structure explicit:
\[
 \mathcal K_P(\mathbf u)
 =\sum_{D,E\in\Sigma_P}\ \sum_{\mathbf y,\mathbf y'}
 \ind{\Gamma_{D,E}}\,w_Dw'_E
 e^{2\pi i(\phi_D-\phi'_E)}.
\]
Here fields of $D$ are evaluated at $(\mathbf x,\mathbf y)$ and primed
fields of $E$ at $(\mathbf x',\mathbf y')$; the two path tuples range
independently over their respective components' bound paths.
The constraint $\Gamma_{D,E}$ requires both guards to hold,
$O_{C,D}=O'_{C,E}=\mathbf c$,
$O_{Q,D}=\mathbf o$, $O'_{Q,E}=\mathbf o'$, and
$\mathsf H_D=\mathsf H'_E$.
Thus, each term pairs two paths with the same hidden history and the
specified outputs, multiplying their real weights and taking the
difference of their phases to account for complex conjugation.

The difference $\Delta(\mathbf u)=\mathcal K_P(\mathbf u)-\mathcal K_Q(\mathbf u)$
has the same guarded-sum form, with contributions from $Q$ subtracted.
\textsc{ReduceKernel} applies the reductions of
Section~\ref{sec:cooperative} before evaluating remaining path sums.
The residual still combines weights and phase factors whose algebraic
relations must be taken into account to determine whether it vanishes.
\emph{Coefficient normalization}, performed by \textsc{EncodeFormula},
collects these contributions in a common basis.

For residuals whose weights and phase factors lie in the supported
algebraic number field $\mathbb K$, coefficient normalization proceeds
as follows. Since $\mathbb K$ is a finite-dimensional vector space over
$\mathbb Q$, choosing a basis $\{\beta_j\}$ gives every element of
$\mathbb K$ a unique representation as a rational linear combination of
these basis elements.
Normalization adds coordinates componentwise and
reduces products using the field's defining relations, collecting all
contributions to each basis element.

Boolean guards and output-selection constraints are retained symbolically
and determine which contributions are present for each index tuple $\mathbf u$.
After all bound sums are eliminated or evaluated, this yields
\begin{equation}
 \Delta(\mathbf u)=\sum_j d_j(\mathbf u)\beta_j,\qquad
 \Delta(\mathbf u)\ne0\ \Longleftrightarrow\
 \bigvee_j d_j(\mathbf u)\ne0.
 \label{eq:residual}
\end{equation}
Here $d_j$ are rational-valued coordinate functions of $\mathbf u$.
Linear independence of
$\{\beta_j\}$ over $\mathbb Q$ therefore implies that
$\Delta(\mathbf u)=0$ exactly when every coordinate is zero.
This criterion applies to the combined coordinates of the complete
residual, not to individual summands. SMT thus checks whether any Boolean
index tuple yields a nonzero rational coordinate, without reasoning
directly about complex algebraic constants.

\paragraph{SMT queries.}
\textsc{EncodeFormula} encodes the coefficient comparison as a
quantifier-free formula $\Phi(\mathbf u,\boldsymbol\xi)$, where $\boldsymbol\xi$
contains encoding auxiliaries, satisfying
\begin{equation}
 \exists\boldsymbol\xi.\ \Phi(\mathbf u,\boldsymbol\xi)
 \quad\Longleftrightarrow\quad
 \bigvee_j d_j(\mathbf u)\ne0.
 \label{eq:encoding-contract}
\end{equation}
An UNSAT result establishes $\Delta\equiv0$ and yields \EQ{}; a SAT result
identifies a differing kernel entry and yields \NEQ{} by
Eq.~\eqref{eq:kernel-equiv}. If the residual
cannot be fully encoded or the solver is inconclusive within the resource
limits, the result is \UNK{}.

\subsection{Reduction}\label{sec:cooperative}
This section presents the reductions underlying \textsc{ReduceHPS} and
\textsc{ReduceKernel} in Algorithm~\ref{alg:checker}.
These reductions serve two complementary objectives.
At the HPS level, reductions aim to preserve and expose structural
correspondence between the two symbolic states, facilitating the
isomorphism-based comparison of Section~\ref{sec:representation}.
At the density-kernel level, reductions extract common factors, cancel
matching contributions, and eliminate reducible bound paths, thereby
simplifying the residual obligations encoded for SMT in
Section~\ref{sec:decision}.
Both levels retain factored expressions and avoid premature expansion of
path sums.

\paragraph{Feedback summaries.}\label{sec:regions}
The basic measurement--correction pattern is
\[
 R\mathrel{:=}\;V;\ c:=\mathsf{measure}\ q;\
 \mathsf{if}\ c\ \mathsf{then}\ V_1\ \mathsf{else}\ V_0,
\]
where $V,V_0,V_1$ are unitary programs (Definition~\ref{def:unitary-program}).
The outcome $c$ is neither read after $R$ nor selected as an observable
output. Such patterns arise in measurement-based
computation and teleportation~\cite{danos2007measurement,gottesman1999teleportation}.

Feedback reduction checks whether the corrected outcomes realize a common
transformation, differing only in input-independent scalar weights.
Let $\Sigma_R$ be the HPS collection for $R$ under the specified
initialization, retaining all outputs needed by the continuation or
selected by $\iota$. As in Section~\ref{sec:observable}, each local history
$\mathbf h$ determines an operator through
$K_{\mathbf h}\ket{\mathbf x}=\ket{\psi_{\Sigma_R,\mathbf h}(\mathbf x)}$,
with unchanged classical coordinates omitted. The history records the
measurement outcome and any discarded outputs.

We seek a single operator $T_{\mathrm{sum}}$ such that
$K_{\mathbf h}=\alpha_{\mathbf h}T_{\mathrm{sum}}$ for every $\mathbf h$
and every symbolic input, where $\alpha_{\mathbf h}\in\mathbb C$ are
input-independent and $\sum_{\mathbf h}|\alpha_{\mathbf h}|^2=1$.
By Eq.~\eqref{eq:observable}, summing the history contributions then gives
$\sum_{\mathbf h}K_{\mathbf h}\rho K_{\mathbf h}^{\dagger}
=T_{\mathrm{sum}}\rho T_{\mathrm{sum}}^{\dagger}$
for every input density operator $\rho$. Thus, $T_{\mathrm{sum}}$ summarizes
$R$ without retaining its local history.
\irene{} constructs $T_{\mathrm{sum}}$ by reducing and aligning the HPS
expressions for corrected outcomes, then merging matching contributions
with their weights combined at the density-operator level.
Replacement requires the criterion above and, for a component subgroup,
disjoint histories from the remaining components before and after replacement.

\paragraph{Graph-based expression reduction.}
On the representations of Section~\ref{sec:representation}, constant
propagation, XOR cancellation, AND idempotence, and common-factor extraction
simplify expressions without distributing nested products.
Local phase expansion exposes cancellations using
$\alpha(B_L\oplus B_R)=\alpha B_L+\alpha B_R-2\alpha B_LB_R$
for Boolean $B_L,B_R$ and real $\alpha$.
Positive Davio decomposition isolates dependence
on a bound path $y$:
\begin{equation}
 B_0=B[y\leftarrow0],\quad B_1=B[y\leftarrow1],\quad
 \delta_yB=B_0\oplus B_1,\qquad B=B_0\oplus(y\land\delta_yB).
 \label{eq:cofactor}
\end{equation}
If all dependence on $y$ occurs in a sign $(-1)^{B(y,\mathbf z)}$, the
remaining contribution $\mathcal A(\mathbf z)$ is independent of $y$, giving
\begin{equation}
 \sum_{y\in\{0,1\}}(-1)^{B(y,\mathbf z)}\mathcal A(\mathbf z)
 =2(-1)^{B_0(\mathbf z)}\ind{\delta_yB(\mathbf z)=0}\mathcal A(\mathbf z).
 \label{eq:fourier}
\end{equation}
Here $\mathcal A$ denotes the complete remaining contribution.
Summing the two signed contributions eliminates $y$ while preserving
measurement histories.

Guard reasoning applies Gaussian elimination over
$\mathrm{GF}(2)$ to XOR constraints~\cite{laitinen2012parity}.
Nonlinear subexpressions serve as formal matrix columns, not independent
Boolean variables. Row XORs expose contradictions or relations that
determine bound paths. A row $0=1$ eliminates the contribution;
a derived relation $y=F(\mathbf z)$, with $F$ independent of bound path
$y$, permits substitution:
\begin{equation}
 \sum_{y\in\{0,1\}}\ind{G(y,\mathbf z)}\mathcal A(y,\mathbf z)
 =\ind{G(F(\mathbf z),\mathbf z)}\mathcal A(F(\mathbf z),\mathbf z).
 \label{eq:unique}
\end{equation}
Substitution introduces no multiplicity factor and applies to every field,
including coefficients and histories. Remaining guards and constraints on
free coordinates are retained.

\paragraph{Path-sum planning.}
Path-sum planning combines dependency-guided decomposition, elimination
ordering, and local expansion to simplify the bound sums in HPS amplitudes
and density kernels. Decomposition identifies independent summation
subproblems by analyzing dependencies in the complete summand; paths within the same
selector or constraint remain coupled.

Suppose the remaining paths partition into disjoint groups
$\mathbf y^{(1)},\ldots,\mathbf y^{(k)}$, the coefficient $C(\mathbf u)$ is
path-independent, and every other factor depends on only one group.
Then the sum decomposes as
\begin{equation}
 \sum_{\mathbf y^{(1)},\ldots,\mathbf y^{(k)}}
 C(\mathbf u)\prod_{i=1}^k \mathcal A_i(\mathbf y^{(i)},\mathbf u)
 =C(\mathbf u)\prod_{i=1}^k
 \left(\sum_{\mathbf y^{(i)}}\mathcal A_i(\mathbf y^{(i)},\mathbf u)\right).
 \label{eq:factorization}
\end{equation}
Each resulting sum can be reduced independently before multiplying the
factors, which may share free coordinates $\mathbf u$.
Common factors remain unexpanded during comparison.

Within each subproblem, bounded search explores elimination orders and
local expansions, evaluating sums over selected bound path variables by
combining their $0$- and $1$-cofactors.

\section{Evaluation}
\label{sec:evaluation}
The evaluation addresses three research questions:
\begin{description}
\item[RQ1: Effectiveness and Efficiency.] How does \irene{} compare with
existing checkers in coverage and runtime across benchmark families?
\item[RQ2: Ablation Study.] How do \irene{}'s individual optimization groups
contribute to its effectiveness and efficiency?
\item[RQ3: Real-World Bug Finding.] Can \irene{} serve as an
equivalence-checking oracle to uncover bugs in quantum compiler optimizations?
\end{description}

\subsection{Experimental Setup}
\label{sec:corpus}
\irene{} is implemented in Rust, accepts bounded OpenQASM~2/3 programs
within its supported fragment~\cite{cross2022openqasm}, and uses Bitwuzla
for residual bit-vector queries~\cite{niemetz2023bitwuzla}.

\paragraph{Benchmarks.}
We evaluate \irene{} on \NumPairs{} program pairs in seven benchmark suites,
whose characteristics are summarized in Table~\ref{tab:benchmarks}.
The pairs cover arbitrary-input and fixed-initialization checks, each with
specified output correspondences. Here, hybrid programs are not purely
unitary and contain measurement (including final readout), reset, or
classical control.
Unitary-gate counts expand source-defined gates and subroutine calls, but retain
standard-library gates.
\begin{table}[t]
\begingroup
\DeclareRobustCommand{\benchAll}{\tikz[baseline=-.55ex]\filldraw[line width=.35pt] (0,0) circle (2.2pt);}
\DeclareRobustCommand{\benchNone}{\tikz[baseline=-.55ex]\draw[line width=.35pt] (0,0) circle (2.2pt);}
\DeclareRobustCommand{\benchSome}{\tikz[baseline=-.55ex]{\fill (0,2.2pt) arc (90:270:2.2pt) -- cycle;\draw[line width=.35pt] (0,0) circle (2.2pt);}}
\caption[Benchmark characteristics.]{Benchmark characteristics. \benchAll, \benchSome, and \benchNone{}
indicate all, some, and no programs of a given type, respectively.
Size entries report P50/P95/mean over per-pair maxima.
Qubit and clbit counts include only bits used by operations or specified interfaces.}
\label{tab:benchmarks}
\centering
\footnotesize
\setlength{\tabcolsep}{2pt}
\renewcommand{\arraystretch}{1.2}
\begin{tabular*}{\linewidth}{@{\extracolsep{\fill}}l*{7}{c}@{}}
\toprule
Benchmark & SQbricks~\cite{ricciardi2025sqbricks} & \shortstack{SQbricks-\\Gen~\cite{ricciardi2025sqbricks}} & \shortstack{Qubit\\Reuse~\cite{decross2023reuse}} & \shortstack{Hybrid-\\QASM~\cite{cross2022openqasm}} & CaQR~\cite{hua2023caqr} & IterTestQ~\cite{paltenghi2026itertestq} & Quokka~\cite{mei2026quokka}\\
\midrule
\# Pairs & 242 & 490 & 10 & 26 & 111 & 180 & 923\\
Unitary & \benchSome & \benchNone & \benchNone & \benchSome & \benchSome & \benchAll & \benchAll\\
Hybrid & \benchSome & \benchAll & \benchAll & \benchSome & \benchSome & \benchNone & \benchNone\\
\midrule
\# Qubits & 33/96/47 & 478/11338/2223 & 8/16/9 & 4/21/6 & 16/16/16 & 11/11/11 & 16/90/34\\
\# Clbits & 28/91/42 & 406/11318/2193 & 8/16/9 & 4/7/4 & 16/16/16 & 0/0/0 & 0/0/0\\
\# Gates & 462/4005/1084 & 1895/52590/10309 & 24/69/31 & 9/29/10 & 395/64283/11292 & 142/359/171 & 445/7410/1656\\
\bottomrule
\end{tabular*}

\endgroup
\end{table}

Expected verdicts come from the benchmark suites. Since tools differ in
their treatment of floating-point precision and numerical tolerances, we
also count a pair as correctly solved when the verdict differs from the
expected label solely because of these differences.
\paragraph{Compared tools.}
\label{sec:baselines}
The baselines are SQbricks~\cite{ricciardi2025sqbricks},
MQT QCEC~\cite{burgholzer2021advanced,burgholzer2022nonunitary},
VeriQC~\cite{hong2022dynamic}, QuPRS~\cite{huang2026quprs}, and
Quokka\#~\cite{mei2026quokka}. These tools cover hybrid lifting,
decision diagrams, tensor representations, and symbolic counting;
Section~\ref{sec:related} discusses their underlying methods.

\paragraph{Environment.}
All experiments use two AMD EPYC 7443 processors
(48 physical cores and 96 hardware threads), running Ubuntu 24.04.4 LTS.
Each program pair is subject to a \TimeLimit{}-second wall-clock limit
and a \MemoryLimit{}\,GiB memory limit per tool--mode run.
QCEC uses two threads per run under the same wall-clock limit.
SQbricks runs separately in parallel (par) and sequential (seq) modes;
we report the best available result for each program pair.

\subsection{RQ1: Effectiveness and Efficiency}
\paragraph{Effectiveness.}
We compare \irene{} with five other equivalence checkers on \NumPairs{}
benchmark program pairs. Table~\ref{tab:main} summarizes the results,
reporting, for each benchmark suite, the number of pairs whose EQ/NEQ
verdicts agree with the ground-truth labels.

\begin{table}[t]
\caption{Equivalence-checking results across benchmark suites.
Each tool column reports solved-pair counts; coverage is the percentage
of all program pairs solved.}
\label{tab:main}
\centering
\begin{minipage}{.94\linewidth}
\small
\setlength{\tabcolsep}{4pt}
\renewcommand{\arraystretch}{1.18}
\begin{tabularx}{\linewidth}{@{}l@{\hspace{9pt}}c@{\hspace{12pt}}*{6}{>{\centering\arraybackslash}X}@{}}
\toprule
Benchmark & \# Pairs & \irene{} & SQbricks & QCEC & VeriQC & QuPRS & Quokka\#\\
\midrule
SQbricks & 242 & \textbf{242} & 169 & 208 & 63 & 31 & 29\\
SQbricks-Gen & 490 & \textbf{490} & 267 & 78 & 34 & 0 & 0\\
Qubit Reuse & 10 & \textbf{10} & 0 & 0 & 0 & 0 & 0\\
Hybrid-QASM & 26 & \textbf{23} & 0 & 0 & 0 & 0 & 0\\
CaQR & 111 & 100 & 0 & \textbf{108} & 82 & 0 & 0\\
IterTestQ & 180 & 177 & 0 & \textbf{180} & 108 & 171 & 94\\
Quokka & 923 & 542 & 0 & \textbf{566} & 257 & 443 & 450\\
\midrule
Overall & \NumPairs{} & \textbf{1584} & 436 & 1140 & 544 & 645 & 573\\
Coverage & & \textbf{79.92\%} & 22.00\% & 57.52\% & 27.45\% & 32.54\% & 28.91\%\\
\bottomrule
\end{tabularx}
\end{minipage}
\end{table}

\irene{} solves \NumDecided{} pairs (\DecisionRate{}), exceeding QCEC by
444 pairs and 22.40 percentage points. It solves 23 of 26 Hybrid-QASM pairs,
whereas no baseline solves a pair in this suite, and all 490 SQbricks-Gen
pairs, compared with 267 for SQbricks and 78 for QCEC.
Hybrid-QASM largely derives from official OpenQASM~3
examples~\cite{cross2022openqasm} and combines measurement-dependent
branches, classical updates, and qubit reuse, including amplitude damping
with environment-qubit reuse. Coverage also reflects source-language and
input/output-mapping support: SQbricks rejects this suite during input
adaptation, while QCEC encounters parsing and runtime errors, including a
deferred-measurement mapping restriction.
\irene{} represents measurement histories and classical control directly
in HPS and uses density-kernel comparison when structural certificates
are insufficient, without first lifting the programs to unitary circuits.

On the unitary Quokka and IterTestQ suites, \irene{} solves 719 pairs
compared with QCEC's 746, including 78 pairs that QCEC does not solve.
Its broader hybrid coverage therefore comes with competitive,
complementary performance on unitary programs.

\begin{figure}[t]
\centering
\includegraphics[width=0.86\linewidth]{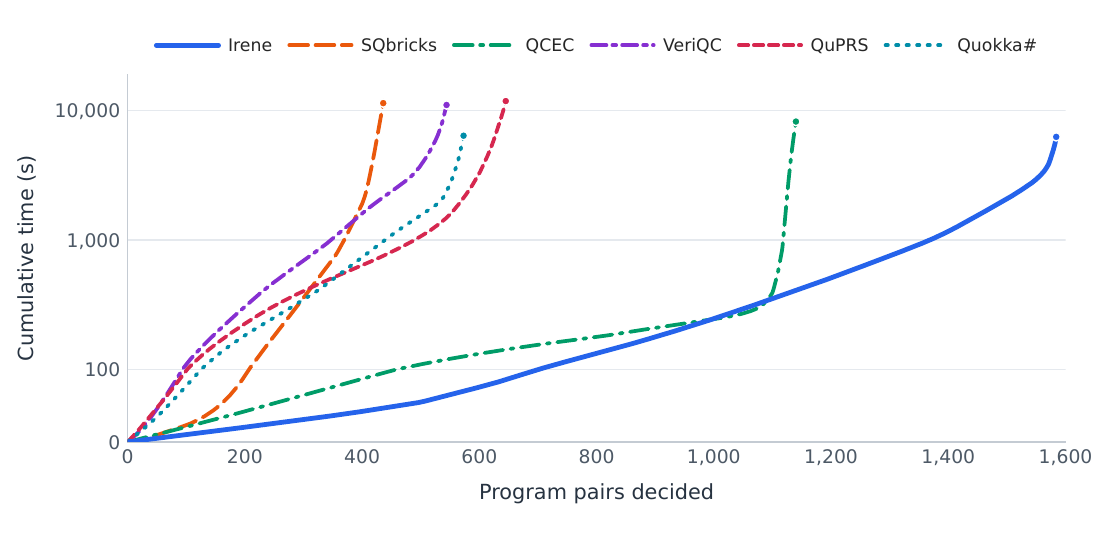}
\caption{Cumulative time for equivalence checking.
Each curve sums the runtimes of solved pairs, ordered from fastest
to slowest. The time axis is linear up to 100\,s and logarithmic above.}
\label{fig:runtime-cactus}
\Description{Six curves show cumulative runtimes for solved pairs under the
evaluation criteria, after sorting each tool's decisions by increasing per-pair runtime.
The endpoints are 1584 pairs for Irene, 436 for SQbricks,
1140 for QCEC, 544 for VeriQC, 645 for QuPRS, and 573 for Quokka sharp.
The vertical coordinate is the sum of runtimes up to each horizontal rank,
on a scale that is linear up to 100 seconds and logarithmic above.}
\end{figure}

\paragraph{Efficiency.}
Figure~\ref{fig:runtime-cactus} shows the relationship between the number of
solved program pairs and cumulative runtime for each tool.
Mean wall-clock time over \irene{}'s 1,584 solved pairs is \MeanTime{} seconds,
with a median of 0.38\,s and a 95th percentile of 13.06\,s.
On common solved pairs, \irene{} is 3.45$\times$, 18.30$\times$,
8.97$\times$, and 5.52$\times$ faster than SQbricks, VeriQC, QuPRS, and
Quokka\#, respectively, over intersections of 436, 523, 619, and 542 pairs.
QCEC is instead 1.72$\times$ faster than \irene{} on their 1,024 common
solved pairs, despite its lower full-corpus coverage. These results suggest
room for further optimization on unitary programs, while demonstrating
\irene{}'s ability to efficiently verify a broader range of hybrid quantum
programs.

\paragraph{Limitations.}
Two repeat-until-success pairs contain unbounded measurement-controlled
loops, outside \irene{}'s bounded model. Most unresolved cases occur in
Quokka, with 254 timeouts,
124 inconclusive results, and three memory-limit failures. All 124
inconclusive cases reach the structural reduction budget and leave residual
kernel-aggregation obligations unresolved. These cases show that local
simplification does not always reduce large unitary computations sufficiently
for the remaining symbolic reasoning to finish within the resource limits.

\subsection{RQ2: Ablation Study}
\begin{figure}[t]
\centering
\includegraphics[width=0.86\linewidth]{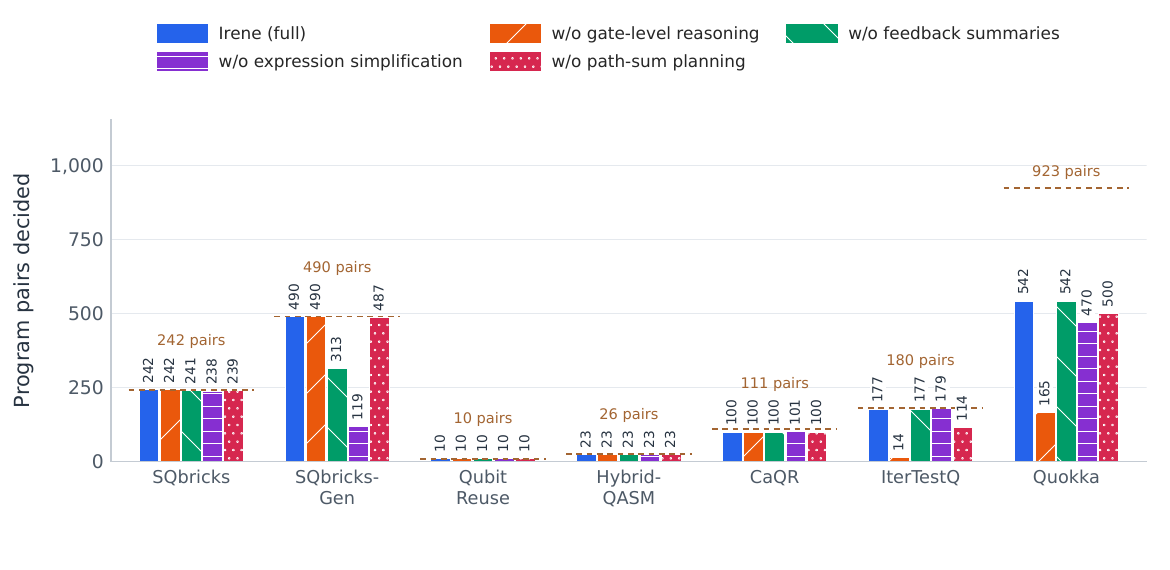}
\caption{Solved program pairs by benchmark suite for the full configuration
and four single-component ablations. Dashed segments indicate the total
number of pairs in each suite.}
\label{fig:ablation-solved}
\Description{Seven groups of five bars compare the full configuration and
four leave-one-out variants. A dashed segment marks each suite's total pair
count. Overall solved counts are 1584 for full Irene, 1044 without gate-level reasoning,
1406 without feedback summaries, 1140 without expression simplification,
and 1473 without path-sum planning.}
\end{figure}

\begin{figure}[t]
\centering
\begin{minipage}[t]{.485\linewidth}
\centering
\includegraphics[width=\linewidth]{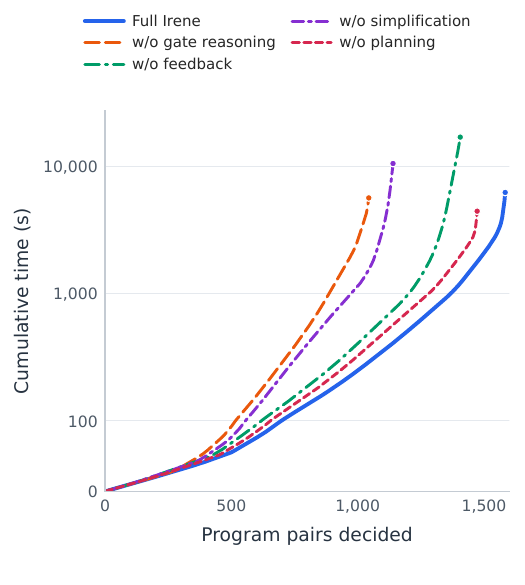}
\caption{Cumulative time for the full configuration and four single-component
ablations. Decided pairs are ordered from fastest to slowest;
the time axis is linear up to 100\,s and logarithmic above.}
\label{fig:ablation-time}
\Description{Cumulative runtimes for five configurations, with a time axis
linear up to 100 seconds and logarithmic above.}
\end{minipage}\hfill
\begin{minipage}[t]{.485\linewidth}
\centering
\includegraphics[width=\linewidth]{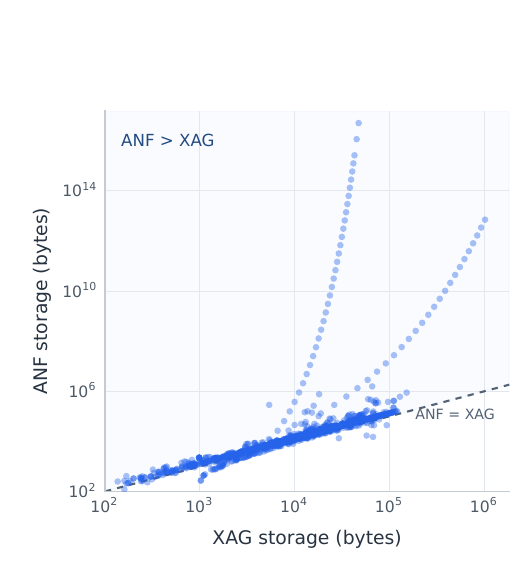}
\caption{XAG versus ANF storage. Each point represents a program pair;
the dashed line marks equal storage.}
\label{fig:representation-scatter}
\Description{Log--log scatter plot
of per-pair peak XAG packed storage versus shared ANF estimated storage,
with independently ranged logarithmic axes and an equal-storage reference
line. Blue circles show 1549 program pairs with exact counts for both representations.}
\end{minipage}
\end{figure}

We compare the full configuration of \irene{} with four leave-one-out
variants, each disabling one optimization group:
\emph{gate-level reasoning} (Section~\ref{sec:gate}),
\emph{feedback summaries} (Section~\ref{sec:cooperative}),
\emph{expression simplification} (Section~\ref{sec:cooperative}), and
\emph{path-sum planning} (Section~\ref{sec:cooperative}).
The gate-level ablation removes rewriting and the associated identity and
trace checks. All variants retain the graph representations and basic
algebraic identities; the expression ablation measures graph simplification,
not the effect of replacing XAGs with expanded polynomials.
Figure~\ref{fig:ablation-solved} groups solved-pair counts by benchmark
suite. Figure~\ref{fig:ablation-time} reports cumulative runtimes.

Overall, every ablation reduces coverage, but the losses differ across
benchmark families. The results support complementary roles for the
optimizations: simplifying unitary computations, reducing measurement
histories, and controlling the cost of residual symbolic reasoning.

Coverage losses from removing gate-level reasoning are confined to the
unitary IterTestQ and Quokka suites. HPS and
density-kernel reasoning alone therefore do not recover the coverage
provided by gate-level certificates. These suites also incur the largest
losses without path-sum planning, even with gate-level reasoning retained.
This supports the progressive design: gate-level checks discharge some
comparisons, while decomposition and elimination ordering remain important
for unresolved path sums.

On measurement-feedback benchmarks, feedback summaries have a more
specialized role: 177 of their 178 lost decisions occur in SQbricks-Gen.
This concentration is consistent with their purpose of replacing local
measurement--correction regions with simpler symbolic summaries, reducing
the histories that subsequent reasoning must retain.

Expression simplification benefits both hybrid and unitary programs, with
the largest coverage losses in SQbricks-Gen and Quokka. Its removal raises
memory-limit failures from three to 79, showing the importance of reduction
beyond the XAG representation itself. Removing gate-level reasoning,
feedback summaries, expression simplification, or path-sum planning
increases geometric mean runtime by 1.60$\times$, 1.51$\times$,
1.54$\times$, and 1.11$\times$, respectively, on each variant's common
solved pairs with full \irene{}.

Simplification does not improve every comparison: disabling expression
simplification loses 468 decisions but gains 24. One possible explanation
is that local rewrites, although semantics-preserving, can obscure common
factors that the existing factor-extraction rules could recognize before
simplification, preventing subsequent cancellation. This suggests a
direction for future work: coordinating local rewrites with factor
extraction and path elimination to preserve structures useful to later
reductions.

\paragraph{Representation size.}
Figure~\ref{fig:representation-scatter} compares per-pair peak XAG and ANF
storage estimates at recorded HPS states. We count XAG nodes and edges
directly and use zero-suppressed decision diagrams~\cite{minato1993zdd} to count ANF monomials
without enumerating them. Both fixed-width storage models account for
sharing: XAGs share subexpressions, whereas the ANF model shares identical
monomials and complete polynomials.
The comparison covers the 1,549 program pairs with exact counts for both
representations. For 30 pairs solved by \irene{}, the estimated explicit ANF storage
exceeds 6\,GiB.
These estimates characterize representation size, not process memory or
the memory requirements of a compressed ANF backend.
\subsection{RQ3: Real-World Bug Finding}
\label{sec:bugs}
LLM-assisted fuzzing generates programs from pass documentation and
function specifications; \irene{} compares each program before and after
transformation. This uncovers two previously unknown bugs in Qiskit, five in Cirq,
two in tket, three in PyZX, and three in PennyLane, totaling
\NumCompilerBugs{} compiler-related bugs.
We also identify two implementation bugs, one each in SQbricks
and HQbricks, the tool implementing HPS~\cite{chareton2026hps}, during our
comparative evaluation. All \NumReportedBugs{}
bugs have been reported. Two Qiskit bugs, one Cirq bug, and the HQbricks bug
have already been confirmed and fixed.
Overall, the compiler-related defects involve lost classical dependencies or
conditions, incorrect gate transformations and circuit extraction, and
phase bookkeeping. The verifier defects concern register-equality
translation in SQbricks and a sign error when combining path amplitudes
in HQbricks. We illustrate these defects with two concrete examples.

In Qiskit, \texttt{BarrierBeforeFinalMeasurements} can reorder measurements
on different qubits that write to the same classical bit, violating their
write-after-write dependency and changing the final classical output even
though the quantum gates are unchanged.
In Cirq, exporting a classically controlled $CCZ$ gate decomposes it into
$H$--$CCX$--$H$, but attaches the condition only to the first instruction.
When the condition is false, the remaining $CCX$ and $H$ still execute.

Standard comparison utilities in these toolchains do not provide general
equivalence checking for hybrid quantum programs: Qiskit's operator
comparison is matrix-based, Cirq's terminal-measurement comparison assumes
measurements occur at the end, and PennyLane's structural comparison does
not establish semantic equivalence~\cite{qiskitOperatorDocs,cirqTerminalEquivalenceDocs,pennylaneEqualDocs}.
By retaining symbolic inputs, quantum coherence, and classical--quantum
correlations induced by measurement and branching, \irene{} detects errors
beyond these comparison capabilities, demonstrating its applicability to
validating compiler optimizations for hybrid quantum programs.

\subsection{Threats to Validity}
\label{sec:threats}
Our symbolic procedure targets exact equivalence, whereas the evaluation
also admits bounded numerical discrepancies. The reported coverage includes
\NumApproxEq{} approximate-equivalence results on purely unitary pairs,
certified by interval arithmetic to have a diamond-norm distance of at most
$10^{-12}$ between their input--output maps. These results do not establish
exact equivalence, and tolerance-based checking of general hybrid quantum
programs is not currently supported.

Aggregate coverage also depends on the benchmark composition and the
program fragments supported by each tool; suite-level results should
therefore be considered alongside the aggregate comparison. The compiler
bugs found are specific to the tested versions, passes, and program-generation
strategy, rather than a comprehensive assessment of these compilers.

\section{Related Work}
\label{sec:related}
\paragraph{Quantum equivalence checkers.}
Decision-diagram and tensor-based methods exploit shared operator structure
and tensor contraction to avoid explicit matrix construction and control
intermediate growth~\cite{burgholzer2021advanced,wei2022sliqec,hong2022tdd,sander2025mpo}.
Standard TDDs decompose tensors by index values and share identical
normalized sub-tensors. When this decomposition produces many distinct
sub-tensors, even a short algebraic description can yield a large decision
diagram; index and contraction orders therefore affect intermediate
representation costs~\cite{hong2022tdd}.
VeriQC extends TDDs to measurements and classical control, but its
scalability still depends on compressing the resulting intermediate
tensors~\cite{hong2022dynamic}.
PBEC avoids full-state representations using local projection constraints
for unitary circuits; its linear scaling at fixed depth depends on
projection locality, which can diminish as depth increases~\cite{yu2025pbec}.
\irene{} instead exploits algebraic relations to eliminate reducible
dependencies through structure-preserving path-sum reduction in
bounded hybrid quantum programs. By removing these dependencies before
constructing larger intermediate representations, it can reduce proof
costs without relying on sub-tensor sharing or depth-bounded projections.

Symbolic and diagrammatic approaches use path-sum or ZX-calculus
rewriting~\cite{amy2019pathsums,kissinger2020pyzx}. Counting-based methods
encode circuit semantics for decision-diagram analysis~\cite{wang2025feynmandd}
or weighted model counting, as in Quokka\#~\cite{mei2026quokka}.
QuPRS combines path-sum reduction with weighted model counting for residual
comparisons~\cite{huang2026quprs}. These algebraic reductions complement
\irene{}'s emphasis on structure-preserving reduction at the HPS level,
where measurement outcomes and their effects on classical control flow
must be handled alongside quantum computation.

For hybrid quantum program equivalence, SQbricks lifts verification to
unitary circuits through deferred measurement, separation, and
projection~\cite{ricciardi2025sqbricks}.
QCEC replaces resets with fresh qubits and defers measurements, converting
classical controls into quantum controls to reuse unitary-circuit
verification~\cite{burgholzer2022nonunitary}.
These reductions can enlarge the circuit by retaining information in
auxiliary qubits; comparing observable behavior also requires accounting
for initialization and for garbage output qubits, as addressed by
partial-equivalence methods~\cite{chen2022partial}.
\irene{} instead symbolically executes bounded hybrid quantum programs and
checks observational equivalence directly, reducing measurement and
control dependencies within HPS without constructing an enlarged unitary
circuit.

\paragraph{Quantum program semantics.}
Path-sum semantics supports algebraic rewriting, composition, and local
summation~\cite{amy2019pathsums,vilmart2021structure,vilmart2023complete,amy2023unbalanced},
as well as deductive verification of circuit-building
programs~\cite{chareton2021qbricks}. HPS provides hybrid-state representations,
rewriting and local reasoning rules, and an assertion language for equivalence
and probabilistic properties, implemented in the HQbricks symbolic execution
engine~\cite{chareton2026hps}.
Building on this framework, \irene{} contributes a comparison-directed
reduction strategy: XAG-based reasoning exposes structural correspondences
and kernel cancellations before expansion can obscure them.
The algorithmic distinction is that reduction is guided by the structural
relations needed for successive equivalence checks.
At the HPS level, retained structure supports graph certificates; at the
kernel level, it enables cancellation before residual expansion.
IHPS instead extends path-sum reasoning to unbounded loops and expected
resource consumption~\cite{chareton2026ihps}.

\paragraph{Compiler verification and testing.}
Verified compilation proves transformation correctness across admissible
inputs~\cite{tao2022giallar,hietala2021voqc}; \irene{} instead validates
individual source--target program pairs.
Static analysis and differential, metamorphic, and fuzz testing detect
defects in quantum programs and their software
stacks~\cite{paltenghi2024lintq,wang2021qdiff,paltenghi2023morphq,iwumbwe2025qutefuzz,paltenghi2026itertestq,xia2026kqfuzz}.
\irene{} complements test generation by serving as an equivalence-checking
oracle for the generated program pairs.

\section{Conclusion}
We presented \irene{}, an observational equivalence checker for bounded hybrid
quantum programs based on structure-preserving symbolic reduction.
It progressively simplifies equivalence obligations at the gate, HPS, and
density-kernel levels, retaining factored symbolic expressions and restricting
expansion to residual comparisons.
Across seven benchmark suites, \irene{} solves \NumDecided{} of
\NumPairs{} program pairs (\DecisionRate{}).
Used as an oracle for LLM-assisted fuzzing, it also uncovers
\NumCompilerBugs{} previously unknown compiler bugs.

\section*{Data Availability}
The \irene{} implementation and reproduction instructions are available
at \url{\ArtifactURL}.

\bibliographystyle{ACM-Reference-Format}
\bibliography{references}
\end{document}